\documentstyle[prl,aps,multicol,epsf]{revtex}
\begin{document}
\draft
\title{Fluctuations and Intrinsic Pinning in Layered Superconductors}
\author{Leon Balents\cite{byline} and David R. Nelson}
\address{Department of Physics, Harvard University, Cambridge, MA 02138}

\date{\today}
\maketitle

\begin{abstract}
A flux liquid can condense into a smectic crystal in a pure layered
superconductors with the magnetic field oriented nearly parallel to
the layers.  If the smectic order is commensurate with the layering,
this crystal is {\sl stable} to point disorder.  By tilting and
adjusting the magnitude of the applied field, both incommensurate and
tilted smectic and crystalline phases are found.  We discuss transport
near the second order smectic freezing transition, and show that
permeation modes lead to a small non--zero resistivity and large but
finite tilt modulus in the smectic crystal.
\end{abstract}

\begin{multicols}{2}
\narrowtext

In the past few years, both
experimental\cite{VGexperiments,BGexperiments}\ and
theoretical\cite{VGtheory,BGtheory}\ work has emphasized the
importance of pinning in type II superconductors.  Although much of
this work has focused on random defects, e.g. twin boundaries, the
layered structure of the copper--oxide materials itself provides a
non--random source of pinning\cite{intrinsicrefs}.  At low
temperatures, since the c-axis coherence length $\xi_{c0} \approx 4 \AA
\lesssim s \approx 12 \AA$, the lattice constant in this direction,
vortex lines oriented in the ab plane are attracted to the regions of
low electron density between the CuO$_2$ layers.  In this paper, we
discuss the phase diagram of intrinsically pinned vortices in near
perfect alignment with the layers.  In contrast to previous work, we
focus on the behavior relatively close to $T_c$, where transport
measurements are most easily performed and hysteretic effects are
weak.  Our research is motivated by the recent experimental work of
Kwok et.  al.\cite{Kwok}, who observed a continuous resistive
transition in ${\rm YBa_2Cu_3O_7}$ for fields very closely aligned
($\theta < 1^\circ$) to the ab plane.

To understand thermal fluctuations, we employ the
boson mapping\cite{NelsonSeung}.  Consider an
isolated flux line oriented along the $a$--$b$ plane.  For $T \gtrsim
80K$, $\xi_c \approx
\xi_{c0}(1-T/T_c)^{-1/2} \gtrsim s$, and the intrinsic pinning barrier
energy per unit length is\cite{BLO}
\begin{equation}
U_p \approx 5\times 10^2 {{\epsilon_0} \over \gamma}\left({{\xi_c}
\over s}\right)^{5/2} e^{-15.8 \xi_c/s},
\label{upinning}
\end{equation}
where the energy scale $\epsilon_0 = (\phi_0/4\pi\lambda_{ab})^2$,
with $\phi_0$ the flux quantum and $\lambda_{ab}$ the $a$--$b$ plane
penetration depth, and where $\gamma \equiv \sqrt{m_c/m_{ab}}$ is the
anisotropy.  With coordinates ${\bf\hat{y}} \parallel {\bf B}
\perp {\bf\hat{c}}$, ${\bf\hat{z}} \parallel {\bf\hat{c}}$, the free
energy of the vortex (described by $x(y)$ and $z(y)$) is
\begin{equation}
F_{\rm v} = \int_0^L \!\!\! dy \!\!\left\{ {{\tilde{\epsilon}_\parallel} \over
2}\left|{{dx} \over {dy}}\right|^2\!\! + {{\tilde{\epsilon}_\perp} \over
2}\left|{{dz} \over {dy}}\right|^2\!\! - U_p \cos 2\pi z/s \right\},
\label{singlevortexfreeenergy}
\end{equation}
where the stiffness constants obtained from anisotropic
Ginzburg--Landau (GL) theory are $\tilde{\epsilon}_\parallel =
\epsilon_0/\gamma$ and $\tilde{\epsilon}_\perp = \epsilon_0\gamma$
(see, e.g. Ref.\onlinecite{BGL}).  
Upon integrating out the $x$--displacement, the
remainder of
Eq.\ref{singlevortexfreeenergy}\ maps to the
Euclidean action of a
quantum particle in a one--dimensional periodic potential.  
In the quantum--mechanical analogy, the particle tunnels between
adjacent minima of the pinning potential, leading to the well--known
completely delocalized Bloch wavefunctions even for {\sl arbitrarily}
strong pinning.  The ``time'' required for this tunneling maps to the
distance, $L_{\rm kink}$, in the $y$--direction between kinks in which
the vortex jumps across one CuO$_2$ layer.  The WKB approximation gives
\begin{equation}
L_{\rm kink} \sim s\sqrt{{{\tilde{\epsilon}_\perp} \over U_p}}
e^{\sqrt{\tilde{\epsilon}_\perp U_p}s/k_{\rm B}T},
\label{WKBresult}
\end{equation}
where $k_{\rm B}T$ plays the role of $\hbar$.  When the sample is
larger than $L_{\rm kink}$ along the field axis, the flux line will
wander as a function of $y$, with
\begin{equation}
\langle [z(y)-z(0)]^2 \rangle \sim Dy,
\label{vortexdiffusion}
\end{equation}
where the ``diffusion constant'' $D \approx s^2/L_{\rm kink}$.  

For $\sqrt{\tilde{\epsilon}_\perp U_p}s \lesssim k_{\rm B}T$, the
pinning is extremely weak, and the WKB approximation is no longer
valid.  Instead, the diffusion constant $D \approx k_{\rm
B}T/\tilde{\epsilon}_\perp$, as obtained from
Eq.\ref{singlevortexfreeenergy}\ with $U_p = 0$.  At much lower
temperatures, when $\xi_c \ll s$, the energy in Eq.\ref{upinning}\
must be replaced by the cost of creating a ``pancake''
vortex\cite{Clem}\ between the CuO$_2$ planes.  In this regime, $L_{\rm
kink} \sim \xi_{ab}(s/\xi_c)^{\epsilon_0 s /k_{\rm B}T}$.

For $T \approx 90K$, as in the experiments of Kwok et. al.\cite{Kwok},
$\xi_c/s \approx 2.3$, and Eq.\ref{upinning}\ gives
$\sqrt{\tilde{\epsilon}_\perp U_p}s/k_{\rm B}T \ll 1$, indicative of
weak pinning and highly entangled vortices in the liquid state.  The
transverse wandering in this {\sl anisotropic} liquid is described by
a boson ``wavefunction'' with support over an elliptical region of
area $k_{\rm B}T
L_y/\sqrt{\tilde{\epsilon}_\parallel\tilde{\epsilon}_\perp}$ with
aspect ratio $\Delta x/\Delta z = \gamma \approx 5$ for ${\rm
YBa_2Cu_3O_7}$ ($L_y \approx 1mm$ is the sample dimension along ${\bf
\hat{y}}$)\cite{BNinprep}.

To explain the observed transition, the interplay between
inter--vortex interactions\cite{intrinsicrefs}\ and thermal
fluctuations must be taken into account in an essential way.  The
experiments of Ref.\onlinecite{Kwok}\ rule out conventional freezing,
which is first order in all known three--dimensional cases.  A
vortex/Bose glass (VG/BG) transition also appears untenable, due to
the purity of the sample (only six twin planes are present, and first
order melting is seen for $\theta > 1^\circ$).  The dynamical scaling
exponents are also inconsistent with VG/BG values.  Instead, we
postulate freezing into an intermediate ``smectic'' phase between
the flux liquid (high temperature) and crystal/glass (low
temperature).  Such smectic freezing, as first discussed by de Gennes
for the nematic--smectic A transition\cite{DeGennes}, can occur via a
continuous transition in three dimensions.  Changes in the applied
magnetic field tune the commensurability of the vortex system with the
underlying periodic structure\cite{Villainreview}.  

\begin{figure}
\epsfysize=1.5truein
\hskip 0.7truein \epsffile{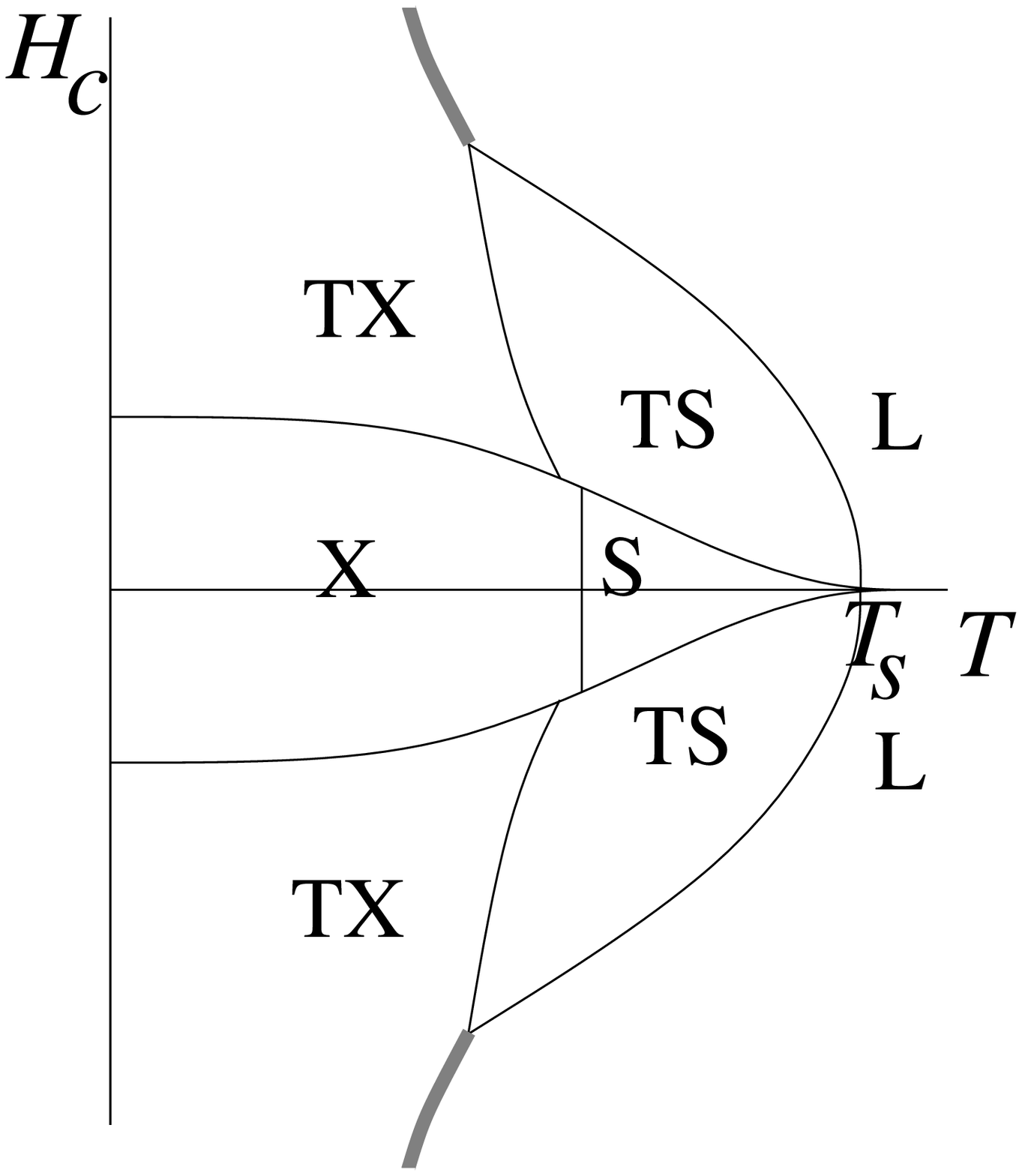}
\caption{Phase diagram for the pure vortex system as a function of
temperature, $T$, and c--axis field, $H_c$.  The fuzzy lines indicate
first order transitions.  A similar phase diagram holds in the $\delta
H_b$--$T$ plane with the TX and TS phases replaced by IX and IS
states, respectively.  The merging of the IX--IS and IS--L lines need
not occur in this case.}
\label{phasediagramfig}\end{figure}

Our analysis leads to the phase diagram shown in
Fig.\ref{phasediagramfig}.  Upon lowering the temperature for $H_c =
0$ and a commensurate value of $H_b$, the vortex liquid (L) freezes
first at $T_s$ into the pinned smectic (S) state, followed by a second
freezing transition at lower temperatures into the true vortex crystal
(X).  When $H_c \neq 0$, tilted smectic (TS) and crystal (TX) phases
also exist.  The TS--L and TX--TL transitions are XY--like, while the
TS--S and TX--X phase boundaries are commensurate--incommensurate
transitions (CITs)\cite{Villainreview}.  At larger tilts, the TX--TS
and TS--L lines merge into a single first order melting line.  As
$H_b$ is changed, incommensurate smectic (IS) and crystal (IX) phases
appear, again separated by CITs from the pinned phases, and an XY
transition between the IS and L states.  Note that the commensurate smectic
order along the $c$ axis is {\sl stable} to point disorder because phonon
excitations are massive (see below).  This stability should {\sl
increase} the range of smectic behavior relative to the (unstable)
crystalline phases when strong point disorder is present.

We now proceed with the derivation of these results; further details
will be given in Ref.\onlinecite{BNinprep}.  
The $a$,$b$, and $c$, crystallographic axes are parameterized by the
coordinates $x$,$y$, and $z$, respectively.  The CuO$_2$ layers are thus
perpendicular to the $z = c$ axis.  The magnetic field is primarily
along the $y = b$ direction.
When
$H_c=0$ and $H_b$ is tuned to commensuration, de Gennes' freezing theory
applies near $T_s$.  Smectic order appears as a density wave,
\begin{equation}
n ({\bf r}) \approx n_0 {\rm Re} \left\{ 1 + \Phi({\bf r})e^{iqz}
\right\},
\label{densitywave}
\end{equation}
where $n_0$ is the background density, and $q = 2\pi/a$ is the
wavevector of the smectic layering (with wavelength $a$).  The complex
translational order parameter $\Phi({\bf r})$ is assumed to vary
slowly in space.  The superconductor is invariant under translations
and inversions in $x$ and $y$, and has a discrete translational
symmetry under $z \rightarrow z + s$, where $s$ is the CuO$_2$
double--layer spacing.  From Eq.\ref{densitywave}, these periodic
translations correspond to the phase shifts $\Phi \rightarrow \Phi
e^{iqs}$.  In the commensurate limit, $a=ms$, with $m$ an arbitrary
integer.  The most general free energy consistent with these
symmetries is
\begin{eqnarray}
F & = \int \! d^3\!{\bf r} \left\{ \right. & {K \over 2}
|(\bbox{\nabla}-i{\bf A})
\Phi|^2 + {r \over 2}|\Phi|^2 + {v \over 4}|\Phi|^4 \nonumber\\
& & \left. - {g \over
2}\left( \Phi^m + \Phi^{*m} \right) + \cdots \right\},
\label{criticalfreeenergy}
\end{eqnarray}
where the coordinates have been rescaled to obtain an isotropic
gradient term.  The ``vector potential'' ${\bf A}$ represents changes
in the applied field $\delta {\bf H} = \delta H_b {\bf\hat{y}} + H_c
{\bf\hat{z}}$, with $A_x = 0$, $A_y = q H_c/H_b$, and $A_z = q\delta
H_b/H_b$.  The form of this coupling follows from the transformation
properties of $\Phi$\cite{DeGennes,BNinprep}.  Additional interactions
with long wavelength fluctuations in the density and tangent
fields\cite{MNhydro} 
are irrelevant to the critical behavior\cite{BNinprep}. 

When $\delta {\bf H} = {\bf A} = 0$, Eq.\ref{criticalfreeenergy}\
is the free energy of an XY model with an $m$--fold symmetry breaking
term.
A second order freezing transition occurs within Landau theory when
$v>0$ and $r \propto T-T_s$ changes sign from positive (in the liquid)
to negative (in the smectic).  The renormalization group (RG) scaling
dimension, $\lambda_m$, of the symmetry breaking term is known {\sl
experimentally} in three dimensions as $\lambda_m \approx 3 - 0.515m -
0.152m(m-1)$\cite{AA}.  For $m>m_c \approx 3.41$, the field $g$ is
irrelevant ($\lambda_m <0$), and the transition is in the XY
universality class.  The magnetic fields used by Kwok
et. al.\cite{Kwok}\ correspond to $m = 9-11$\cite{AGL}, well into this
regime.  The static critical behavior is characterized by the
correlation length exponent $\nu \approx 0.671 \pm 0.005$ and
anomalous dimension $\eta \approx 0.040 \pm 0.003$\cite{exponentrefs}.

Deep in the ordered phase ($r<0$), amplitude fluctuations of $\Phi$
are frozen out.  Writing $\Phi = \sqrt{|r|/v} e^{2\pi i u/a}$,
Eq.\ref{criticalfreeenergy}\ becomes, up to an additive constant,
\begin{equation}
F_{\rm smectic} = \int \! d^3\!{\bf r} \left\{ {\kappa \over 2}
(\bbox{\nabla}u \!-\! {\bf\cal A})^2 - \tilde{g}\cos 2\pi u/s \right\},
\label{smecticfreeenergy}
\end{equation}
where $\kappa = 4\pi^2|r|K/a^2v$, $\tilde{g} = g(|r|/v)^{m/2}$, and
the reduced vector potential is ${\bf\cal A} = {\bf A}/q$.  The
displacement field $u$ describes the deviations of the smectic layers
from their uniform state.  The sine--Gordon term is an effective
periodic potential acting on these layers.  As is well known from the
study of the roughening transition\cite{roughening}, such a
perturbation is always relevant in three dimensions.  The smectic
state is thus {\sl pinned} at long distances (i.e. the displacements
$u$ are localized in a single minima of the
cosine)\cite{decoupledfoot}.

At lower temperatures, provided point disorder is negligible, the
vortices will order along the $x$ axis as well\cite{Feinberg}.  This
is once again a freezing transition at a single wavevector, and is
described by a Landau theory like Eq.\ref{criticalfreeenergy}\ (with
$g=0$ since the modulating effect of the underlying crystal lattice is
much weaker in the $x$ direction), with XY critical behavior.

Next, we consider the effects of ${\bf \delta H} \neq 0$.  When the
fields are small, ${\bf\cal A}$ is an {\sl irrelevant} operator in the
smectic phase (as can be easily seen by replacing the periodic
potential by a ``mass'' term $\propto u^2$).  This implies that the
smectic layers do not tilt under weak applied fields, i.e. $\partial
\langle \partial_y u \rangle/\partial H_c|_{H_c =0} = 0$.
The full susceptibility $c_{44}^{-1} \equiv \partial \langle
B_c \rangle/\partial H_c|_{H_c = 0}$ is obtained from the expression
\begin{equation}
B_c = {{Kq} \over H_b}{\rm Im}\left(\Phi^*\partial_y\Phi\right) +
\left(c_{44,0}^{-1} - r'' |\Phi|^2\right)H_c,
\label{transversefield}
\end{equation}
where $c_{44,0}$ is the tilt modulus obtained from anisotropic GL
theory (without accounting for the discreteness of the layers) and
$r'' \equiv \partial^2 r/\partial H_c^2|_{H_c = 0}$.
Eq.\ref{transversefield}\ has a simple physical interpretation.  The
first term is the contribution to $B_c$ from tilting of the layers
(described by a phase rotation of $\Phi$).  Transverse field
penetration at fixed layer orientation contributes via the second
term.  Such motion arises microscopically from a non--zero equilibrium
concentration of vortices with large kinks extending between
neighboring smectic layers.  Eq.\ref{transversefield}\ predicts a
non--divergent singularity $c_{44}(T) - c_{44}(T_s) \sim
|T-T_s|^{1-\alpha}$ at the critical point, where $\alpha$ is the
specific heat exponent.  At low temperatures in the smectic phase this
crosses over to the much larger value $c_{44} \approx
[\sqrt{\tilde{\epsilon}_\perp/U_p}k_{\rm B}T/(B\phi_0 ms)]
\exp(E_{\rm lk}/k_{\rm B}T)$, where the energy of a large kink is
$E_{\rm lk} \approx \sqrt{\tilde{\epsilon}_\perp U_p} ms$.

As is well known from the study of the sine--Gordon
model\cite{Villainreview}, a larger incommensurability can be
compensated for by energetically favorable ``solitons'', or walls
across which $u
\rightarrow u + s$.  Solitons begin to proliferate when their field
energy per unit area $\sigma_{\rm field} \sim -\kappa{\cal A}s$
exceeds their cost at zero field, $\sigma_0
\sim \sqrt{\kappa\tilde{g}} s$ (estimated from
Eq.\ref{smecticfreeenergy}).  

Physically, these solitons correspond to extra/missing flux line layers
and walls of aligned ``jogs'' for $\delta{\bf H}$ along
the $b$ and $c$ axes, respectively.  In the
former case, this leads to an incommensurate smectic (IS) phase, whose
periodicity is no longer a simple multiple of $s$.  For $\delta{\bf H}
\parallel {\bf\hat{z}}$, the solitons induce an additional periodicity
along the $y$ axis.  This tilted smectic (TS) phase has long range
translational order in two directions\cite{crystalnote}.  The
analogous tilted {\sl crystal} (TX) phase is qualitatively similar,
but has long range order in 3 directions.

Unlike the corresponding CIT in two--dimensional adsorbed
monolayers\cite{Villainreview}, we find that energetic interactions
between widely separated solitons dominate over entropic
contributions\cite{BNinprep}.  The free energy density in the
incommensurate phases is thus
\begin{equation}
f_{\rm soliton} \sim -{{|\sigma|} \over l} + {\Delta \over
l}e^{-l/w},
\label{solitonfreeenergy}
\end{equation}
where $\sigma \equiv \sigma_{\rm field}+\sigma_0 <0$ is the total areal free
energy of the soliton; $\Delta$ and $w$ set the energy and length scales of the soliton
interactions.  Minimizing Eq.\ref{solitonfreeenergy}\ gives a soliton separation $l
\sim w\ln(\Delta/|\sigma|)$ near the CIT.

As the temperature is increased within the IS or TS phases, the system
melts into the liquid.  The IS--L and TS--L transitions are described
by Eq.\ref{criticalfreeenergy}\ with $g=0$ (as can be seen by
reversing the dilation discussed earlier and discarding oscillatory
contributions to $F$ proportional to $g$), and are thus XY--like.

The shape of the CIT phase boundary is of particular experimental
interest.  In the mean field regime, this is obtained from the
condition $\sigma=0$ as $\delta H \sim |r|^\Upsilon$, with
$\Upsilon_{\rm MF} = (m-2)/4$.  By the usual Ginzburg
criterion, mean field theory breaks down for $|r|
\lesssim (k_{\rm B}Tv/K^{3/2})^2$.  To determine the shape of the
 phase boundary in this critical regime, we follow the RG flows
until $|r|$ is order one.  Then $\delta
H_{\rm R} \sim \xi^{\lambda_{\rm H}}\delta H$ and $g_{\rm R} \sim
\xi^{\lambda_m}g$, with $\xi \sim |r|^{-\nu}$.  Rotational invariance
at the rescaled fixed point ($g=0$) implies that the field exponent is {\sl
exactly} $\lambda_{\rm H} = 1$\cite{BNinprep}.  Using these
renormalized quantities, we find $\Upsilon_{\rm crit.} = (|\lambda_m|
+ 2)\nu/2 \approx 4.9 - 7.2$ for the fields used in Ref.\onlinecite{Kwok}.
The IS--L and TX--L phase boundaries are non--singular and are
determined locally by the smooth $\delta {\bf H}$
dependence of $r$.

We next discuss transport measurements using a coarse--grained
approach similar to flux line hydrodynamics\cite{MNhydro}.  The time
evolution of the fluxon density $n({\bf r},t)$ is determined by the
continuity equation
\begin{equation}
\partial_t n + \bbox{\nabla}_\perp\cdot{\bf j}_v = 0,
\label{continuity}
\end{equation}
where $\bbox{\nabla}_\perp = (\partial_x,\partial_z)$, and ${\bf j}_v$ is
the vortex current, determined by the constitutive equation,\cite{MNhydro}
\begin{equation}
\Gamma {\bf j}_v = -n\bbox{\nabla}_\perp{{\delta F} \over
{\delta n}} + n\partial_y{{\delta F} \over {\delta\bbox{\tau}}} 
-\tau_\alpha\bbox{\nabla}_\perp{{\delta F} \over {\delta\tau_\alpha}} + n{\bf f},
\label{constitutive}
\end{equation}
where $\bbox{\tau}$ is the coarse--grained tangent density, $\Gamma$
is related to the Bardeen--Stephen friction coefficient $\gamma_{\rm
BS} = n_0\Gamma$, and the driving force is
\begin{equation}
{\bf f} = {{\phi_0} \over c}{\bf J \wedge \hat{y}} +
\bbox{\eta}.
\label{forces}
\end{equation}
Here ${\bf J}$ is the applied transport current density and
$\bbox{\eta}({\bf r})$ is a random thermal noise.  Upon projecting the
critical smectic density modes near $\pm q{\bf\hat{z}}$ out of
Eq.\ref{continuity}, we find\cite{tangentnote}
\begin{equation}
\gamma_{\rm BS}\partial_t \Phi = - 4q^2{{\delta F_{\rm crit.}} 
\over {\delta\Phi^*}}
- i\mu J_x \Phi - \tilde{\eta},
\label{criticaldynamics}
\end{equation}
where $\mu = q\phi_0 n_0/c$ and $\tilde{\eta}({\bf k}) = in_0 q
\eta_z(q{\bf\hat{z}} + {\bf k})$.  Remarkably, Eq.\ref{criticaldynamics}\
has the same form as the model E dynamics\cite{HalperinHohenberg}\ for
the complex ``superfluid'' order parameter $\Phi$, where now $J_x$ plays the
role of the ``electric field'' in the Josephson coupling.  The actual
electric field is ${\cal E}_x = j_{v,z}\phi_0/c$, leading via
Eq.\ref{constitutive}\ to
\begin{eqnarray}
{\cal E}_x & \approx & -{{n_0\phi_0} \over
{2qc}}{\rm Im}\left(\Phi^* \partial_t\Phi\right)
\nonumber \\
& & + (1 -
|\Phi|^2/2)\left({B \over {H_{c2}}}\right)\rho_{xx,n}J_x,
\label{vortexcurrent}
\end{eqnarray}
where $\rho_{xx,n}$ is the normal state resistivity in the $x$
direction, whose appearance in the last term follows from the relation
$(n_0\phi_0/c)^2/\gamma_{\rm BS} \approx (B/H_{c2})\rho_{xx,n}$.
Eq.\ref{vortexcurrent}\ is interpreted in close analogy with
Eq.\ref{transversefield}.  The
first term is the contribution to vortex flow from translation of the
layers, while motion of equilibrium vortex kinks yields the second
term.  Such flow at ``constant structure'' is analogous to
the permeation mode in smectic liquid crystals\cite{DeGennes}.

The presence of this defective motion implies a small but
non--zero resistivity at the L--S transition.  Near $T_s$,
Eq.\ref{vortexcurrent}\ predicts a singular decrease of the form
$\rho_{xx}(T) - \rho_{xx}(T_s) \sim |T_s - T|^{1-\alpha}$.  At lower
temperatures (but still within the S phase) transport occurs via two
channels.  The permeation mode gives an exponentially small linear
resistivity $\rho_{xx} \sim \exp(-E_{\rm lk}/k_{\rm B}T)$ (above $T_s$,
single layer kinks give $\rho_{xx} \sim \exp(-E_{\rm k}/k_{\rm B}T)$,
with $E_{\rm k} \approx E_{\rm lk}/m$).  Non--linear
transport occurs via thermally activated liberation of vortex
droplets, inside which $u$ (or $u_z$ in the crystal phase) is shifted
by $s$.  Balancing the soliton energy on the boundary with the Lorentz
energy in the interior gives ${\cal E}_{nl} \sim e^{-(J_c/J)^2}$, where
$J_c \sim (c/B)(\kappa\tilde{g})^{3/4}(s/k_{\rm B}T)^{1/2}$.  
A more detailed discussion of the full
scaling form of ${\cal E}(J)$ will be given in Ref.\onlinecite{BNinprep}.

In the TS phase, net vortex motion along the $c$ axis occurs by
sliding soliton walls along the $b$ direction.  The resulting electric
field is proportional to $J$ and the soliton density, leading to an
additional contribution to the resistivity which vanishes at the CIT
like $\rho_{xx}^{\rm soliton} \sim \rho_0^{\rm soliton}/\ln(\Delta/|\sigma|)$.
The situation in the IS phase is more complex, due to the possibility
of a roughening transition for a single soliton wall, and will be
deferred to Ref.\onlinecite{BNinprep}.

Lastly, we consider the effects of weak point disorder, which enters
the free energy as a random field $F_d = \int \! d^3\!{\bf r} n_0
V_d({\bf r}) {\rm Re} \{ \Phi({\bf r})e^{iqz} \}$, where $V_d({\bf
r})$ is a quenched random white noise potential.  Such a perturbation
alters the universality classes of the phase transitions in
Fig.\ref{phasediagramfig}, and renders all but the L and S phases
glassy\cite{BNinprep}.  The stability of the S phase to randomness
(which takes the form $F_d = \int \! d^3{\bf r} 2n_0\sqrt{|r|/v}
V_d({\bf r})\cos 2\pi(u+z)/a$ in this regime) is a consequence of the
phonon ``mass'' $\tilde{g}/s^2$.  The nature of the glassy phases and
altered critical behavior will be discussed in
Ref.\onlinecite{BNinprep}.

It is a pleasure to acknowledge discussions with Daniel Fisher,
Matthew Fisher, Randall Kamien, and Onuttom Narayan.  This research
was supported by the NSF through Harvard University's Materials
Research Lab and through grant DMR--91--15491.  L.B.'s
work was also supported by the NSF at MIT by grants DMR--93--03667 and
PYI/DMR--89--58061 and at Harvard by grant DMR--91--06237.

\end{multicols}


\begin{references}
\bibitem[*]{byline} Also at Physics Department, Massachusetts Institute
of Technology, Cambridge, MA, 02139.

\bibitem{VGexperiments} R. H. Koch et. al., Phys. Rev. Lett. {\bf 63}, 1511
(1989); P. L. Gammel et. al., Phys. Rev.  Lett. {\bf 66}, 953 (1991).

\bibitem{BGexperiments} L. Civale et. al., Phys. Rev.  Lett. {\bf 67},
648 (1991); 

\bibitem{VGtheory} D. S. Fisher, M. P. A. Fisher, and D. A. Huse,
Phys. Rev. {\bf B43}, 130 (1991).

\bibitem{BGtheory} D. R. Nelson and V. M. Vinokur, Phys. Rev.
B{\bf 48}, 13060 (1993).

\bibitem{intrinsicrefs} B. I. Ivlev and N. B. Kopnin, J. Low Temp.
Phys. {\bf 80}, 161 (1990); B. I. Ivlev, N. B. Kopnin, and V. L.
Pokrovsky, J. Low Temp. Phys. {\bf 80}, 187 (1990).

\bibitem{Kwok} W. K. Kwok et. al., Phys. Rev. Lett. {\bf 72}, 1088 (1994).

\bibitem{NelsonSeung} D. R. Nelson, Phys. Rev. Lett. {\bf 60}, 1415
(1988); D. R. Nelson and S. Seung, Phys. Rev. B{\bf 39}, 9153 (1989).

\bibitem{BLO} A. Barone, A. I. Larkin, and Yu. N. Ovchinnikov, J.
Supercond. {\bf 3}, 155 (1990).

\bibitem{BGL} G. Blatter, V. B. Geshkenbein, and A. I. Larkin, Phys.
Rev. Lett. {\bf 68}, 875 (1992).



\bibitem{Clem} J. R. Clem, Phys. Rev. B{\bf 43}, 7837 (1991).

\bibitem{BNinprep} L. Balents and D. R. Nelson, in preparation.

\bibitem{DeGennes} P. G. de Gennes and J. Prost, {\em The Physics of
Liquid Crystals,} (Oxford University Press, New York, 1993).

\bibitem{Villainreview} J. Villain, in {\it Ordering in Strongly
Fluctuating Condensed Matter Systems}, edited by T. Riste (Plenum, New
York, 1980), p. 221.

\bibitem{MNhydro} M. C. Marchetti and D. R. Nelson, Physica C {\bf
174}, 40 (1991); and references therein.

\bibitem{AA} A. Aharony et. al., Phys. Rev. Lett. {\bf 57}, 1012 (1986).

\bibitem{AGL}See, e.g. L. J. Campbell, M. M. Doria, and V. G. Kogan,
Phys. Rev. B{\bf 38}, 2439 (1988).

\bibitem{exponentrefs} J. Zinn--Justin, {\em Quantum Field Theory and
Critical Phenomena}, (Oxford Univ. Press, New York, 1990), p. 619.

\bibitem{roughening} J. D. Weeks, in {\em Ordering in Strongly
Fluctuating Condensed Matter Systems,} edited by T. Riste (Plenum, New
York, 1980), p. 293.


\bibitem{decoupledfoot}  Nevertheless, a description as a system of
weakly coupled two dimensional vortex liquids fails, due to
the neglect of dislocation pairs in neighboring layers\cite{BNinprep}.

\bibitem{Feinberg} More complex structures are possible.  See
D. Feinberg and A. M. Ettouhami, Physica Scripta {\bf T49}, 159 (1993).



\bibitem{crystalnote} Without energetically forbidden terminations of
soliton walls, a stack of layers in three dimensions is automatically
crystalline.

\bibitem{tangentnote} As in model C dynamics\cite{HalperinHohenberg},
dynamical fluctuations in the long wavelength tangent and density
fields are irrelevant since $\alpha <0$\cite{BNinprep}.

\bibitem{HalperinHohenberg} P. C. Hohenberg  and B. I. Halperin, Rev.
Mod. Phys. {\bf 49}, 435 (1977).




\end{references}
\end{document}